\documentstyle[aps,prl,epsfig] {revtex}
\begin{document}

\title{ Dynamical Quasi-Stationary States in a system 
with long-range forces}

\author{V. Latora\footnote{E-mail: vito.latora@ct.infn.it}
and {A. Rapisarda\footnote{E-mail: andrea.rapisarda@ct.infn.it}}}

\address{Dipartimento di Fisica e Astronomia,  Universit\'a di Catania,
\\ and INFN sezione di Catania,
Corso Italia 57 I 95129 Catania, Italy}

\date{\today}
\maketitle

\begin{abstract}
The Hamiltonian Mean Field model describes a system of N fully-coupled 
particles showing a second-order phase transition as a function of the 
energy. The dynamics of the model presents interesting features in a 
small energy region below the critical point. 
In particular, when the particles are prepared in a ``water bag'' 
initial state, the relaxation to equilibrium is very slow.  
In the transient time the system lives in a dynamical quasi-stationary 
state and exhibits anomalous (enhanced) diffusion and L\'evy walks. 
In this paper we study temperature and velocity distribution  
of the quasi-stationary state and we show that the 
lifetime of such a state increases with N. In particular when 
the $N\rightarrow \infty$ limit is taken before the $t \to \infty$ 
limit, the 
results obtained are different from the expected canonical predictions. 
This scenario seems to confirm a recent conjecture proposed by C.Tsallis.
\end{abstract}
\bigskip

\section{Introduction}
The Hamiltonian Mean Field (HMF) model describes N classical 
particles moving on the unit circle and interacting through an 
infinite range potential ~\cite{ruffo}. 
HMF has been recently studied both analitycally and 
numerically~\cite{latora98,latora99,prl99,pro2000,ruffod}. 
The model has the advantage of having an exact solution 
in the canonical ensemble and therefore allows to study  
microscopic dynamics in connection to  thermodynamic
macroscopic features.
The analytical calculation in the canonical ensemble predicts 
a second-order phase transition from a clustered phase to a 
``gaseous'' one,  where the particles are homogeneously 
distributed on the circle. 
A microcanonical solution, recently obtained, 
has been discussed by S.Ruffo at this conference~\cite{ruffod}.
The numerical simulations show that the dynamics is chaotic 
and the Lyapunov exponents are maximal at the critical point. 
In the energy range close to the critical point   
the relaxation to equilibrium is very slow, though the dynamics 
is strongly chaotic.  
In particular when the system is started in out-of-equilibrium initial 
conditions it shows the presence of Quasi-Stationary States (QSS) whose 
relaxation time increases with the size of the system.  
The particle motion is superdiffusive 
in the transient quasi-stationary regime preceeding equilibration.

In the following sections we remind the reader the details
of the model and then we focus on the quasi-stationary states 
discussing superdiffusion, velocity distributions and lifetimes.   
As a main result we show that close to the critical point, if the continuum 
($N\rightarrow \infty$) limit is performed before the $t \to \infty$ 
limit, canonical equilibrium is never reached and quasi-stationary
superdiffusive states live forever.

\section{The model}

The Hamiltonian of our  system is :  
\begin{equation}
        H(\theta,p)=K+V ~,
\end{equation}
where 
\begin{equation}
       K= \sum_{i=1}^N  {{p_i}^2 \over 2} ~~,~~~~~ 
       V= {1\over{2N}} \sum_{i,j=1}^N  [1-cos(\theta_i -\theta_j)]
\end{equation}
\noindent
are the kinetic and potential energy.  
The system consists of $N$ classical particles moving on the unit circle: 
each particle is characterized by the angles $\theta_i$ 
and the conjugate momenta $p_i$, and interacts with all the others. 
One can also see the model in a different way: 
if we consider  a spin vector associated to each particle
${\bf m}_i=[cos(\theta_i), sin(\theta_i)]$, 
the Hamiltonian then describes a linear chain of 
$N$ classical fully-coupled spins, similarly to the XY model.  
With this interpretation we can define a 
total magnetization ${\bf M}={\frac{1}{N}}\sum_{i=1}^N {\bf m}_i$. Thus 
the system is a ferromagnet at low
energy~~ and shows a second-order phase transition ~~at the critical 
energy density~~ $U_c=0.75$  $ (U=E/N)$, corresponding to the critical
temperature $T_c=0.5$ in the canonical 
ensemble ~\cite{ruffo,latora98,latora99}. On the other hand,
 one gets an antiferromagnetic behavior
 by changing the sign of the interaction. This case  has been
studied in detail too ~\cite{ruffo,pro2000,ruffo2000}.

Considering the components of the magnetization vector 
${\bf M}=(M_x,M_y)$ and expressing the potential in the 
following way  
\begin{equation} 
V= { N\over 2} (1- M_x^2+M_y^2) = { N\over 2} (1- M^2) ~~,~~~ 
\label{eqmoto1} 
\end{equation}
the equations of motion for the $N$ particles are 
\begin{equation} 
\dot{\theta_i}={p_i}~~, ~ ~ ~ ~ ~~~\dot{p_i}  = -sin(\theta_i) M_x  + cos(\theta_i) M_y  
~~,~~~ 
i=1,...,N~~.
\label{eqmoto2} 
\end{equation}
The latter equations can be also recast in the form 
\begin{equation} 
\dot{\theta_i}={p_i}~~, ~ ~ ~ ~ ~~~\dot{p_i}  = - M sin(\theta_i - \phi ) ~~,~~~ 
i=1,...,N~~~,
\label{eqmoto} 
\end{equation}
where $(M,\phi)$ are respectively the modulus and the phase 
of the total magnetization vector $\bf M $. 
These equations are formally equivalent to those of
a perturbed pendulum.                                   
The equations of motion  were integrated numerically 
by means of a 4th order  simplectic  algorithm
~\cite{yoshida}  which allowed, using a  time step $\delta t=0.2$, very long 
integration times (the  average number of steps was of the order of
$10^{7}$)  with a relative error in the total conserved energy  smaller than 
$\Delta E/E=10^{-5}$ 
(the details can be found in Refs~\cite{latora98,latora99}). 
Such long integration time is necessary in order to reach equilibration 
when out-of-equilibrium initial conditions are used. In the following 
we present numerical simulations for systems with different $N$ values 
and energies $U=E/N$. In particular we focus on the region below the 
critical point.

\section{L\'evy walks, Superdiffusion and Quasi-Stationary States}

In this section we discuss the transport properties of single particles  
in the transient out-of-equilibrium regime for $0.6 \le U < U_c$ 
and we study the characteristics  
of the quasi-stationary macroscopic states.
Diffusion and transport of a particle in a medium or in a fluid 
flow are characterized by the mean square displacement $\sigma^2(t)$,
which in the long-time limit, is given by the equation
\begin{equation}
\label{anoma}
    \sigma^2(t) \sim  t^{\alpha}~,
\end{equation}
\noindent
with $\alpha=1$ for normal diffusion.  When $\alpha \ne 1$  
one has anomalous 
diffusion , and in particular 
{\em subdiffusion} if $0<\alpha<1$ and 
{\em superdiffusion} if  $1<\alpha<2$. 
Anomalous diffusion has been mainly studied in chaotic systems with  
only a few  degrees of freedom \cite{levypro,kla}, and only recently 
in high-dimensional systems \cite{kaneko,torc}.
In order to study the non-equilibrium properties of HMF model, 
we start our system in a ``water bag'', 
i.e. an initial condition obtained by 
putting all the particles at $\theta_i=0$ and
giving them a uniform distribution of momenta with a
finite width centered around zero. By following the 
dynamics of each particle, we compute the variance 
in the angle $\theta$ according to the expression
\begin{equation}
    \sigma^2(t) = < (\theta - <\theta> )^2 >~,
\end{equation}
where  $< ~.~>$  indicates  the average over the $N$ particles, 
and we fit the value of the exponent $\alpha$ in Eq. (\ref{anoma}). 
In fig.~1 we show the typical motion of a particle
as a function of time for a system with $N=1000$ and $U=0.69$. 
In (a) we report the time evolution of the angle and in (b) the respective
momentum time evolution. 
It is clearly visible that the particle is sometimes trapped
by the main cluster (formed by all the  other particles) and oscillates 
around it with zero average momentum. But the
particle can also frequently experience free walks
in angle, with an average  momentum rather constant and greater 
than that of the cluster. 
This dynamics, typical in this range of energy and timescale, 
gives rise to a variance which is not linear in time. 
In the inset (c) we plot the time evolution of the mean 
square displacement  (averaged over all the particles).
We have checked, by fitting the numerical data in different time intervals, 
that the slope is $\alpha=1.45\pm0.1$ and we report in the figure 
a straight  dotted line with this slope for comparison. 
Superdiffusion becomes normal only after  
a cross-over time and one recovers the slope $\alpha= 1$ only 
at equilibrium (see ref.~\cite{prl99} for details).

\begin{figure}
\begin{center}
\epsfig{figure=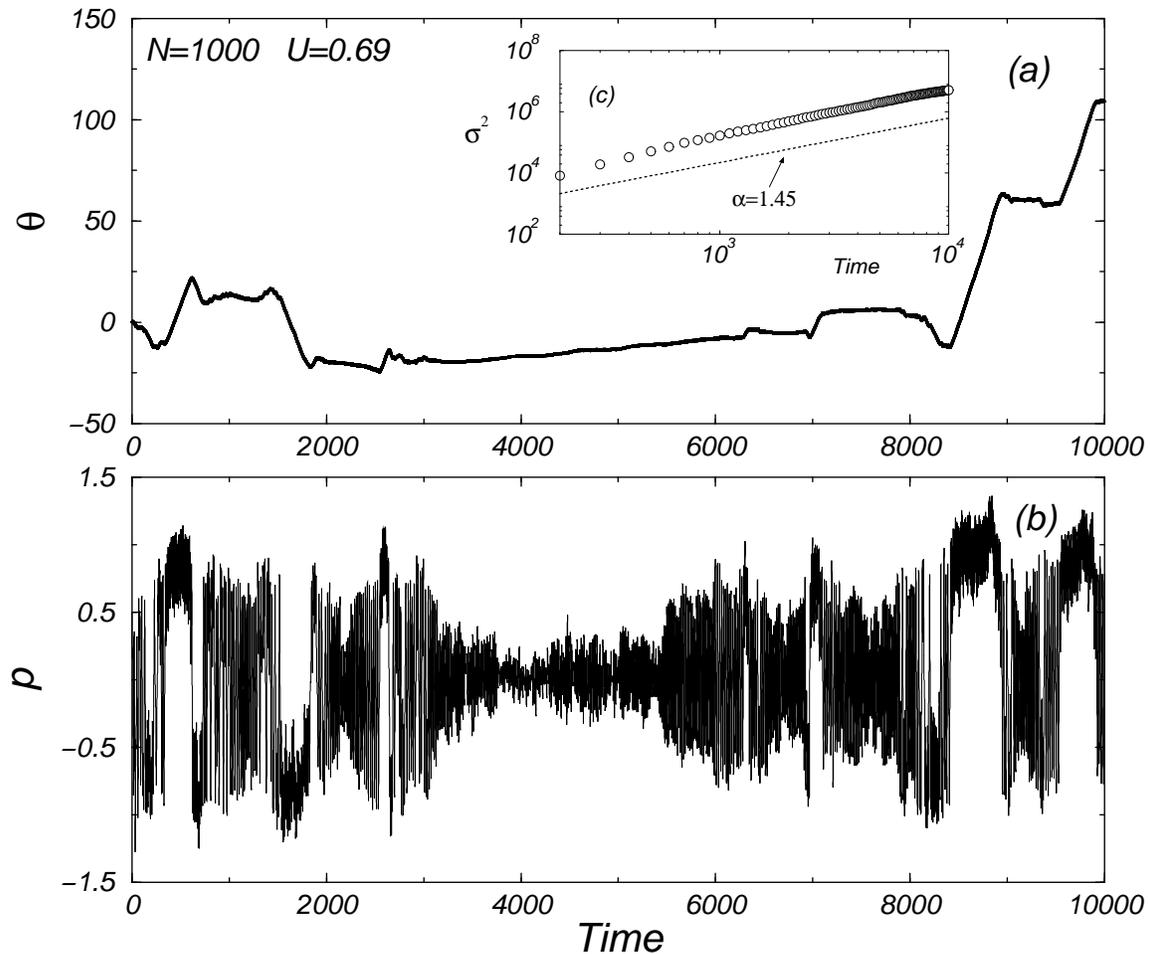,width=13truecm,angle=-90}
\end{center}
\caption{ Typical superdiffusion of a single particle in the 
transient out-of-equilibrium regime for N=1000 and U=0.69. We show
the time  evolution of the angle in (a) and of the corresponding 
momentum (b).  In the inset (c) we show
the behavior of the variance vs time (open symbols) and a straight 
dotted  line with slope $\alpha=1.45$. See text for further details.}
\end{figure}

This erratic non-brownian behavior  can be better 
characterized by calculating the trapping times  and walking times 
  probability distributions.
This study has been discussed in detail in ref.  \cite{prl99},  where it 
has been shown that, using the model 
by Klafter and collaborators   \cite{kla,barkai},  one gets 
a consistent scenario of L\'evy walks and superdiffusion . In fact 
one obtains    trapping times and walking times  probability  
distributions which are power laws, i.e. 
\begin{equation}
    P_{walk}(t) \sim t^{-\mu}
~,~~~~
    P_{trap}(t) \sim t^{-\nu}
~,
\end{equation}
with fitted numerical values for U=0.69 equal to
$\mu=2.14~,~\nu=1.58$. These exponents, according to the Klafter 
and Zumofen model, obey to the formula  
\begin{equation}
    \alpha= 2 +  \nu -\mu~,
\end{equation}
which gives a value  $\alpha=1.44~$, consistent with the  one
extracted from the fit of the numerical time evolution  for the mean square 
displacement, shown in the inset of fig.1.
It has also been checked that in a region close to the critical point, 
from U=0.6 to U=0.75, these values do not to depend in a 
sensitive way on the system size and only slightly on the energy density U.
One has therefore L\'evy walks and superdiffusive behavior, which however 
turns again into normal transport after a cross-over time.
In Ref.~\cite{prl99} we have shown that this cross-over time 
coincides with the equilibration one.

In order to better study the non-equilibrium properties of our model,
  we show, in fig.~2, the time evolution of the temperature calculated through 
the average kinetic energy ($T=2<K>/N$).  We report in particular 
the case U=0.69 for  
different $N$ values. The curves are the result of the averaging over ten different
runs. The figure shows that  
the microcanonical temperature converges to a well defined   plateau, before relaxing to the 
canonical (greater) value, also indicated. The complete relaxation is shown only for N=1000,
while the simulation was truncated for the other cases.
This persistent and constant non-equilibrium value of the temperature, shown in the figure, 
indicates   the presence of a dynamical  Quasi-Stationary
State (QSS).  
The simulations clearly show that the lifetime of this QSS 
(length of the plateaus) increases with N, and that the value of the saturation 
temperature converges in the continuum limit,  to a particular value 
$T_{QSS}$. The latter 
 lies on the continuation
of the homogeneous canonical phase, which represents an unstable branch 
in the microcanonical 
ensemble ~\cite{ruffod}.  We show in the inset of the figure  the comparison
between the canonical caloric curve (full curve) and the plateau temperature 
for N=20000 (open circles).
In fig.~3 we compare the behavior of the QSS temperature as a function of N 
(open circles) with  the canonical value of the temperature (dashed line). 
The difference between the two temperatures increases 
with the size of the system and reaches a saturation value around N=20000.  
The maximal size adopted in the numerical simulations was N=25000. 
The  results shown in figures 2 and 3 indicate a  divergence 
of the QSS lifetime in the limit $N\rightarrow \infty$ limit 
and a convergence of the microcanonical temperature to a
value lower than the canonical one . In other words,
in the continuum limit the system does not  relax to the standard canonical equilibrium 
and remains forever in the QSS with a temperature $T_{QSS}$.  
The nature of these states seems to be purely dynamical and connected to a particular 
initialization. Further numerical investigation is in progress 
in this respect. From the analytical point of view,
it has been shown by  Ruffo at this conference, that these states correspond to a local 
minimum of the free energy ~\cite{ruffod}.

\begin{figure}
\begin{center}
\epsfig{figure=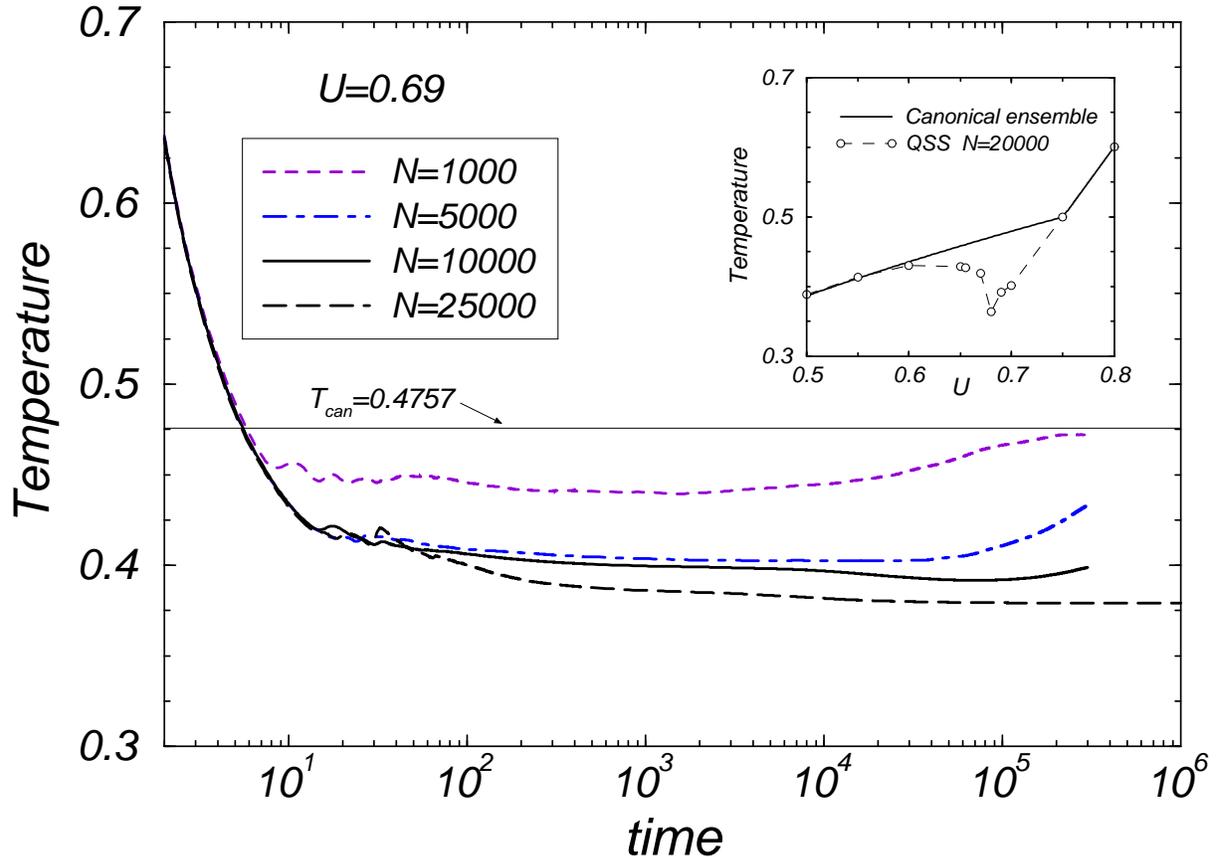,width=16truecm,angle=0}
\end{center}
\caption{ Time evolution of the temperature for different systems sizes.
Each curve is the result of an  average over ten runs.
A plateau, which increases with N exists before equilibration to the canonical ensemble.
The latter is shown only for N=1000.
 We show in the inset the equilibrium caloric curve (full curve) and the 
out-of-equilibrium one (open symbols) 
calculated for N=20000. The latter 
corresponds to the value of the  dynamical plateaus for different energies.}
\end{figure}

\begin{figure}
\begin{center}
\epsfig{figure=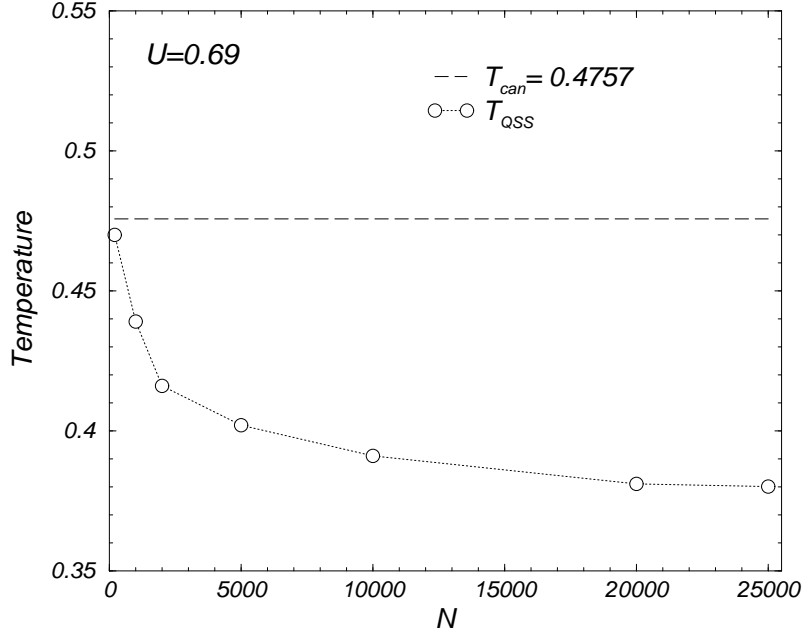,width=9truecm,angle=-90}
\end{center}
\caption{ Temperature of the Quasi Stationary State at U=0.69  
vs N (open circles) in comparison
with the canonical value $T_{can}=0.4757$  (dashed line).}
\end{figure}

\begin{figure}
\begin{center}
\epsfig{figure=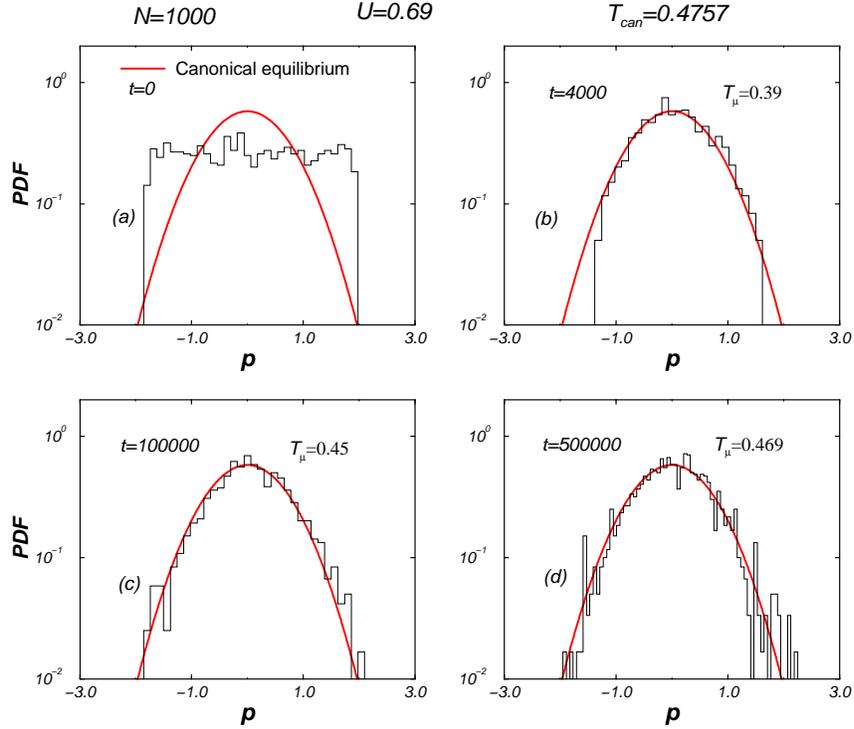,width=10truecm,angle=-90}
\end{center}
\caption{ The evolution of Probability Distribution Function (PDF) 
for the momentum at different timescales $t$ and for U=0.69  (histogram) 
is compared with the theoretical  one (full curve).  
Water bag initial conditions are used. The microcanonical 
temperatures $T_\mu$ is  reported at each time for comparison. The numerical 
simulation shown corresponds only to one run.}
\end{figure}

Finally, we show in fig.4  the time evolution of the Probability Distribution 
Function (PDF)  for the momentum (histogram) at four different timescales.
The case shown refers to $U=0.69$ and $N=1000$. 
Though the numerical distribution fuction is constructed  
by means of only one event and therefore is not a perfectly smooth curve, 
the transient regime (see in particular panel (b) at time 
$t=4000$  and compare also with fig.2) shows results 
clearly different from the gausssian function   (reported as a full curve)
\begin{equation}
\label{gauss}
    F(p) =   {{1}\over \sqrt{2 \pi T}} ~  e^{-p^2 / T}
\end{equation}
predicted by the Vlasov solution and by the canonical ensemble 
\cite{latora98,latora99}. 
The logarithmic scale  shows that the 
tails of the numerical histograms are missing  in the transient regime,  fig.4(b), 
producing a microcanonical temperature $T_{\mu}$  (reported in the figures for each time) 
smaller than the canonical one $T_{can}=0.4757$. The latter 
 is finally reached only after a very long integration time, i.e. $t=500000$.
Preliminary and more refined results (not reported here) 
show a power-law decay in the tails of the QSS velocity 
distributions, at variance with the exponential 
behavior predicted by the canonical equilibrium. 
Summarizing,  it seems that by performing the limit 
$N\rightarrow \infty$ before that one  $t \to \infty$ 
the system, initialized in a water bag, will stay indefinitely out of equilibrium and 
the momentum distribution will never develop a Maxwellian curve.
This fact seems to support  the idea of C. Tsallis who suggested
 his generalized thermostatistics 
as the appropriate formalism  
 to describe  such dynamical Quasi-Stationary 
States~\cite{tsallis}.

\section{Discussion and Conclusions}

We have presented new numerical studies on the metastable Quasi-Stationary States
recently found in the dynamics of the Hamiltonian Mean Field model. 
HMF is an useful model to study the links between dynamics and 
thermodynamics in a system with long-range forces.
In fact the model can be solved exactly in the canonical ensemble and
this solution can be compared with microcanonical dynamical simulations 
for different sizes of the system. 

When the microcanonical simulations are started in a out-of-equilibrium 
initial state, for example the so-called ``water bag'', the results show a 
clear indication of the presence of QSS, in a 
transient temporal regime before relaxation.  We have shown that 
the temperature of QSS reaches a well defined value in the continuum limit 
and that the lifetime of these states increases with N. 
Moreover the velocity distributions of QSS do not show Maxwellian tails.  
The corresponding caloric curve lies below the canonical one, showing a well 
defined backbending. The latter simulates a first order phase 
transition, at variance with the second order one obtained at equilibrium. 
Superdiffusion and L\'evy walks are present in this transient 
out-of-equilibrium regime, for energies close to the critical one, 
implying a coexistence of a liquid (clustered particles) and a gas 
(free particles) phase.
It has also been found that the cross-over time 
from anomalous to normal diffusion coincides
with the relaxation time.  The fact that the
ralaxation time  diverges with N, though the dynamics is strongly 
chaotic in this region, independently on the size,
is rather counter-intuitive and still not fully
understood. 
This  behavior is probably due to the fact that the greater the number 
of  particles considered in the system, 
the stronger are the correlations in the dynamics. 
Close to the critical point,  when the initial big cluster (in a  
water bag,  all the particles 
are on top of each other at initial time) 
tries to fragment into smaller
clusters, relaxation is probably hindered 
by these smaller fragments, which try to capture 
the free particles and form dynamical barriers. 
This effect of course increases with N. 
 
Finally we would like to stress the similarity of the scenario 
indicated by our simulations with the conjecture 
by C.Tsallis of a different equilibrium for non-extensive systems.  
In fact, the HMF model, due to the long-range nature of the interaction,   
has a non-extensive character and shows strong correlations 
in space and time: the phase space is probably a 
multifractal in the transient regime. It is  a fascinating challenge 
left for future investigations 
to study in detail 
the eventual connection between the  out-of-equilibrium dynamics in HMF 
(and other similar models \cite{celia,giansanti}) 
and the generalized nonextensive statistics discussed 
at  this conference. 
It would be also important to study the link  to real experimental systems 
\cite{delca,solomon}.    

This paper  is part of a work in progress in  collaboration with S. Ruffo and C. Tsallis.

\end{document}